\documentclass[10pt,final,journal,twocolumn]{IEEEtran}
\ifCLASSINFOpdf
\else
\fi
\usepackage{cite}
\usepackage[cmex10]{amsmath}
\usepackage{amsthm}
\usepackage{amsmath}
\usepackage{bm}
\usepackage{algorithmic}
\usepackage{stfloats}
\usepackage{multicol,multienum}
\usepackage{graphicx}
\usepackage{color}
\usepackage{amssymb}
\usepackage{float}
\usepackage{amsfonts}
\usepackage[compatibility=false]{caption} 
\usepackage{subcaption}
\usepackage{ltablex}        

\usepackage{booktabs}      
\usepackage[table]{xcolor} 
\usepackage{tabularx}      
\usepackage{array}         
\usepackage{multicol}      
\usepackage{caption}       
\definecolor{tbBg}{RGB}{235,244,252}   
\definecolor{oddBg}{RGB}{250,250,250}  
\definecolor{evenBg}{RGB}{255,255,255} 
\definecolor{grid}{RGB}{200,200,200}   

\usepackage{makecell}
\arrayrulecolor{grid}
\newcolumntype{C}[1]{>{\centering\arraybackslash}m{#1}}  

\usepackage{array} 


\begin{document}  
\title{Integrated Sensing and Covert Communication In Low-Altitude Networks: A Smart Radio Environment Perspective}
\author{
Jianyu Wei, Haichao Wang, Laixian Peng, Jiangchun Gu, Ziqi Liu, Lifeng Chen and Guoru Ding


\thanks{This work was supported in part by the National Natural Science Foundation of China ( No. 62271501, No. 62401626 and No. 61871400) and in part by the Jiangsu Province Natural Science Foundation under Grant  (BK. 20240200). (Corresponding author: Haichao Wang and Laixian Peng.)}

\thanks{Jianyu Wei, Haichao Wang, Laixian Peng, Jiangchun Gu, and Guoru Ding are with the College of Communications Engineering, PLA Army Engineering University, Nanjing, 210007, China (email: wjianyu@aeu.edu.cn; lxpeng@aeu.edu.cn; whcwl0919@sina.com;  gujiangchungjc@sina.com; 3011745933@qq.com; guoru$\_$ding@yeah.net).

Ziqi Liu is with the Jiangsu Provincial Meteorological Observatory, Nanjing, 210007, China. (email: dz1928005@smail.nju.edu.cn)

}

}

\IEEEpeerreviewmaketitle
\maketitle
\begin{abstract}
The rise of low-altitude economies and 6G is driving the evolution of low-altitude networks (LANs), making communication security a pressing concern. Unlike traditional security approaches, covert communication offers enhanced protection by hiding the transmission behavior itself. Integrated sensing and communication (ISAC), a key technology of 6G, efficiently supports both sensing and communication tasks through hardware integration, thereby promising significant gains for covert communication. Nevertheless, the complexity and dynamics of urban environments pose critical challenges. Drawing on the latest advances in smart radio environment (SRE) technologies, this paper introduces them into integrated sensing and covert communication (ISACC) to suppress covert channel fading and counteract sensing precision loss in LANs. We first survey the applications and state-of-the-art findings of ISACC in LANs, highlighting key practical challenges. Subsequently, we introduce the core concept of SRE and elaborate on its enabling techniques across four dimensions. To deliver more insights, we explore potential pathways for integrating SRE into ISACC. To maximize covert throughput, a reinforcement learning-based case study is conducted by jointly optimizing flight trajectory, jamming power, movable antenna position, bandwidth allocation, and beamforming vectors. Simulation results show that the proposed scheme achieves superior performance compared to the benchmark. Finally, some open challenges and potential directions are discussed.

\end{abstract}

\IEEEpeerreviewmaketitle

\section{Introduction}

\IEEEPARstart{T}HE arrival of the 6G era and the rapid rise of low-altitude economies are opening up vast markets for unmanned aerial vehicle (UAV)-based applications. Integrated sensing and communication (ISAC), a key enabling technology for 6G, promises to deliver both high-throughput, ultra-reliable, low-latency communication and precise sensing and positioning for low-altitude networks (LANs) [1]. These capabilities provide strong technical support for diverse use cases, such as logistics delivery, autonomous driving, and smart transportation [1]. Moreover, leveraging their high maneuverability and line-of-sight (LoS) link advantages, UAVs can further extend the performance boundaries and coverage of communication and sensing when integrated with ISAC. However, the openness and broadcast nature of wireless channels also elevate the risks of signal interception and malicious attacks, thereby raising considerable concerns about communication security in LANs. Covert communication, also known as low probability of intercept communication [2], can establish covert links for UAV communications, thereby significantly enhancing transmission security. In addition, when equipped with ISAC, UAVs can acquire real-time position information of target nodes, which in turn enhances covert performance. Recently, integrated sensing and covert communication (ISACC) in LANs has attracted considerable attention [3].

The integrated design of ISAC noy only lowers hardware complexity and facilitates deployment on UAVs, but also enables real-time warden localization through sensing signals, directly solving the problem of locating a non-cooperative adversary. Meanwhile, the sensing signal can also serve as a jamming mechanism to disrupt the warden’s detection. As a result, ISACC has emerged as a hot research topic. In complex urban environments, large-scale and small-scale fading often coexist. While UAVs can preserve LoS links through attitude adjustments, avoiding non-line-of-sight conditions becomes challenging when serving multiple users. These channel degradations not only undermine covert performance but also offer natural cover for the warden to conceal its location, further complicating covert communication and sensing. Hence, new solutions are urgently needed to improve UAV covert communication in complex urban environments.

\begin{figure*}[!htb]   
\centering
\includegraphics[width=190mm,height=130mm]{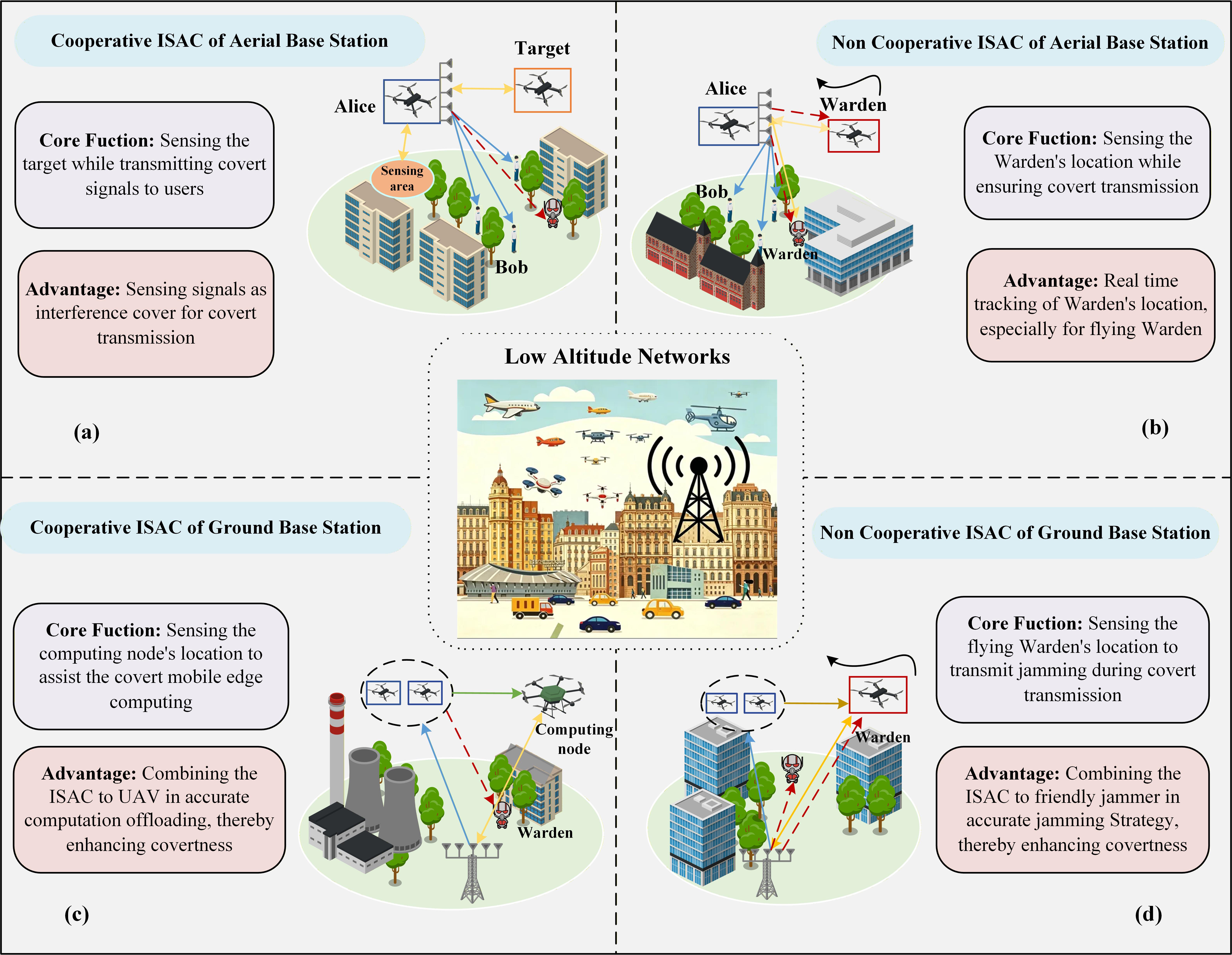}
\caption{The typical application scenarios and advantages of ISACC in LANs: (a) Cooperative ISAC of aerial base station; (b) Non-cooperative ISAC of aerial base station; (c) Cooperative ISAC of ground base station; (d) Non-cooperative ISAC of ground base station.}
\label{Fig1}
\end{figure*}

Advances in electromagnetic information theory have opened up new possibilities for tackling this challenge through smart radio environment (SRE) technologies [4]. In SRE, the wireless channels could be continuously manipulated to improve their performance and coverage [4]. Initially, SRE was closely linked to reconfigurable intelligent surface (RIS) in the electromagnetic dimension. The RIS consists of multiple low-power passive reflecting elements, each capable of actively reconfiguring the wireless propagation environment by adjusting the phase or amplitude of reflected signals [5]. Introducing RIS can not only effectively mitigate covert channel fading in complex urban environments but also further suppress signal leakage toward the warden, thereby significantly improving the covertness of UAV communications. With the integration of emerging technologies, SRE could flexibly manipulate wireless channels across multiple dimensions. Centered on SRE, this paper investigates how to jointly enhance UAV sensing and covert transmission in LANs. We begin by surveying the state-of-the-art of ISACC in LANs, and highlight key practical challenges. Next, we summarize SRE technologies from four dimensions: electromagnetic, frequency, spatial, and waveform. Moreover, we explore promising application scenarios of SRE for ISACC in complex LANs, along with their performance advantages. Based on this, we present a case study of movable antenna (MA)-enabled ISACC, conforming that SRE significantly improve UAV covert performance. Finally, we conclude by discussing the key research challenges and promising future directions for SRE-enabled ISACC in LANs.

\section{Overview Of integrated sensing and covert communication in LANs}

This section presents typical application scenarios of ISACC in LANs, highlighting their core functions and key benefits. Then, we further review the current research and identify existing bottlenecks and challenges.

\subsection{The state-of-the-art findings of ISACC in LANs}

Low-altitude airspace typically refers to the region within 1,000 meters above ground [6]. As key carriers of LANs, UAVs require a high level of communication covertness for missions including emergency rescue, urban counter-terrorism, and privacy protection. Thus, we focus on UAV covert communication in LANs [6]. Given the widespread adoption of ISAC in such scenarios, as illustrated in Fig. 1, we analyze its application from two perspectives: aerial base stations (ABSs) and ground base stations (GBSs). In ABS scenarios, although UAVs could extend the coverage of covert communications, their LoS links can also be vulnerable to malicious attacks. ISAC offers a promising way to enhance UAV covert communication, primarily through two sensing modes ---- cooperative and non-cooperative.

\textbf{Cooperative ISAC of aerial base station (CISAC-ABS):} ISAC devices deployed on UAVs serving as ABSs can simultaneously sense friendly targets and provide covert communication services to ground users. This mode is particularly common in low-altitude urban smart transportation scenarios. The sensing and covert communication signals exhibit a mutually beneficial relationship: Sensing signals not only detect moving friendly targets in real time but can also jam the warden, thereby protecting UAV covert transmissions. Meanwhile, communication signals can deliver the acquired target information to ground users, enabling them to track target locations.

\textbf{Non-cooperative ISAC of aerial base station (NCISAC-ABS):} This mode is especially relevant in low-altitude urban emergency rescue missions where GBSs are compromised. In such scenarios, ISAC devices can simultaneously perform downlink covert transmission while detecting the real-time location of the aerial warden. This allows the communication party to obtain the warden's location and take timely actions such as avoidance or jamming, thereby further enhancing covert performance.

Due to the limited onboard energy and the diversity of heterogeneous nodes in LANs, UAV swarms often rely on GBSs to provide different services such as communication, computing, and navigation. ISAC can further enhance both covert communication and mobile computing tasks for UAV swarms, primarily through two sensing modes ---- cooperative and non-cooperative.

\textbf{Cooperative ISAC of ground base station (CISAC-GBS):} This mode is typically applied in scenarios such as smart transportation and urban air mobility. Here, ISAC devices installed on GBSs can both sense the mobile computing center and use communication signals to deliver precise location to UAV swarms. This approach not only helps each UAV achieve better covert computation offloading performance, but also cuts down each UAV’s energy consumption, thereby improving the overall energy efficiency of the covert edge computing system.

\textbf{Non-cooperative ISAC of ground base station (NCISAC-GBS):} This mode is commonly used in urban counter-terrorism where aerial wardens are present. In such cases, GBSs equipped with ISAC devices can perceive the real-time location of the aerial warden and share it with the UAV swarm. Then, the jammer can impose jamming signal on the warden and conduct its expulsion. This cooperative mechanism not only mitigates the effects of jamming on the own receiver, but also effectively disrupts the warden's detection, thereby ensuring covert computation offloading tasks within the UAV swarm.


\begin{table*}[t]
\centering
\small
\setlength{\tabcolsep}{3.5pt}
\renewcommand{\arraystretch}{1.5}
\caption{A comparative summary of current research on ISACC in LANs}
\begin{tabular}{
  >{\centering\arraybackslash\columncolor{tbBg}}m{1.3cm}    
  >{\centering\arraybackslash\columncolor{tbBg}}m{1.9cm}    
  >{\centering\arraybackslash\columncolor{tbBg}}m{1.4cm}    
  >{\centering\arraybackslash\columncolor{tbBg}}m{1.8cm}    
  >{\centering\arraybackslash\columncolor{tbBg}}m{4cm}    
  >{\centering\arraybackslash\columncolor{tbBg}}m{2.6cm}    
}
\toprule[1pt]
\rowcolor{tbBg}
\textbf{Ref.} &
\textbf{Application style} &
\textbf{Types of A2G links} &
\textbf{Role of UAVs} &
\textbf{Main Feature} &
\textbf{Covertness}\\
\midrule

\rowcolor{oddBg}
\lbrack3\rbrack & CISAC-ABS & LoS link & Transmitter & First introduce ISAC into the UAV covert communication & Channel uncertainty  \\
\rowcolor{oddBg}
\lbrack5\rbrack & NCISAC-ABS & LoS link & Relay & Introduce RIS into ISAC-assisted UAV covert communication & Channel uncertainty \\
\lbrack7\rbrack & NCISAC-ABS & Probability LoS Link & Warden \& Transmitter & Consider the ISAC-assisted UAV covert and secure communication & Noise uncertainty and warden’s position uncertainty  \\
\lbrack8\rbrack & NCISAC-BBS & LoS link & Warden & Multiple jammers are introduced to jointly jam the flying warden & Channel uncertainty \\  
\rowcolor{evenBg}
\lbrack9\rbrack & CISAC-BBS & LoS link & Warden \& Transmitter & First integrating the sensing, communication and computing in UAV covert communication & Noise uncertainty  \\
\rowcolor{oddBg}
\lbrack10\rbrack& CISAC-BBS & LoS link & Receiver \&
Jammer & Introduce the full-deplex UAV to receive data and jam the warden & Jamming power uncertainty \\
\bottomrule[1pt]
\end{tabular}
\end{table*}

\subsection{Research Status and Open Problems}

ISACC in LANs has recently attracted significant research interest. Table 1 summarizes the main research efforts in this field. [3,5,7] investigate the application about ABSs of ISACC in LANs. Specifically, [3] pioneers the ISAC into UAV covert communication. To tackle the dynamic of urban environments, [5] explores the covertness gains brought by RIS. Moreover, [7] employs ISAC to jointly address both covertness and secrecy of UAV communications. [8-10] focus on related applications about GBSs of ISACC in LANs. Specifically, [8] and [10] integrate ISAC into UAV covert communication under active jamming strategy and friendly jamming strategy, respectively. [9] pioneers the study of covertness in integrated sensing, communication, and computing systems. Despite these diverse efforts, several critical problems still persist of ISACC in LANs:

\textbf{Simplistic air-ground channel modeling:} Most current research relies on the LoS assumption for air-ground links, which breaks down in complex urban LANs for practical operations. However, channel fading has a negative impact on both covert transmission and target sensing. Consequently, exploring ISACC under realistic channel fading in complex LANs is a critical and pressing challenge.

\textbf{UAV flight jitter issues:} Mounting RIS on a UAV is an innovative and interesting idea [2]. Nevertheless, unavoidable attitude jitter during actual UAV flight can significantly impair the signal reflection of the RIS. Hence, seeking more practical and robust engineering solution has emerged as a pressing research priority.

\textbf{Limited sources of uncertainty:} Exploiting uncertainty to disturb the warden's detection is necessary for UAV covert communication [6]. So far, current research has largely focused on channel, noise, and warden location uncertainties. Exploring novel uncertainty dimensions will be critical for advancing covert communication in future LANs.

\section{Realization of SER-Enabled ISACC in LANs }

In complex urban environments, the sensing and covert transmission performance of UAVs can be significantly degraded by channel fading effects. SRE technologies, such as RIS, actively reconfigure the electromagnetic environment to effectively mitigate these adverse effects. It is worth noting that RIS is not the sole enabler for SRE. Therefore, this section first systematically analyzes SRE technologies across four dimensions: electromagnetic, frequency, spatial, and waveform. Subsequently, we explore the potential application scenarios of SRE in ISACC tasks.

\begin{figure}[!htb]
\centering
\includegraphics[width=0.5\textwidth]{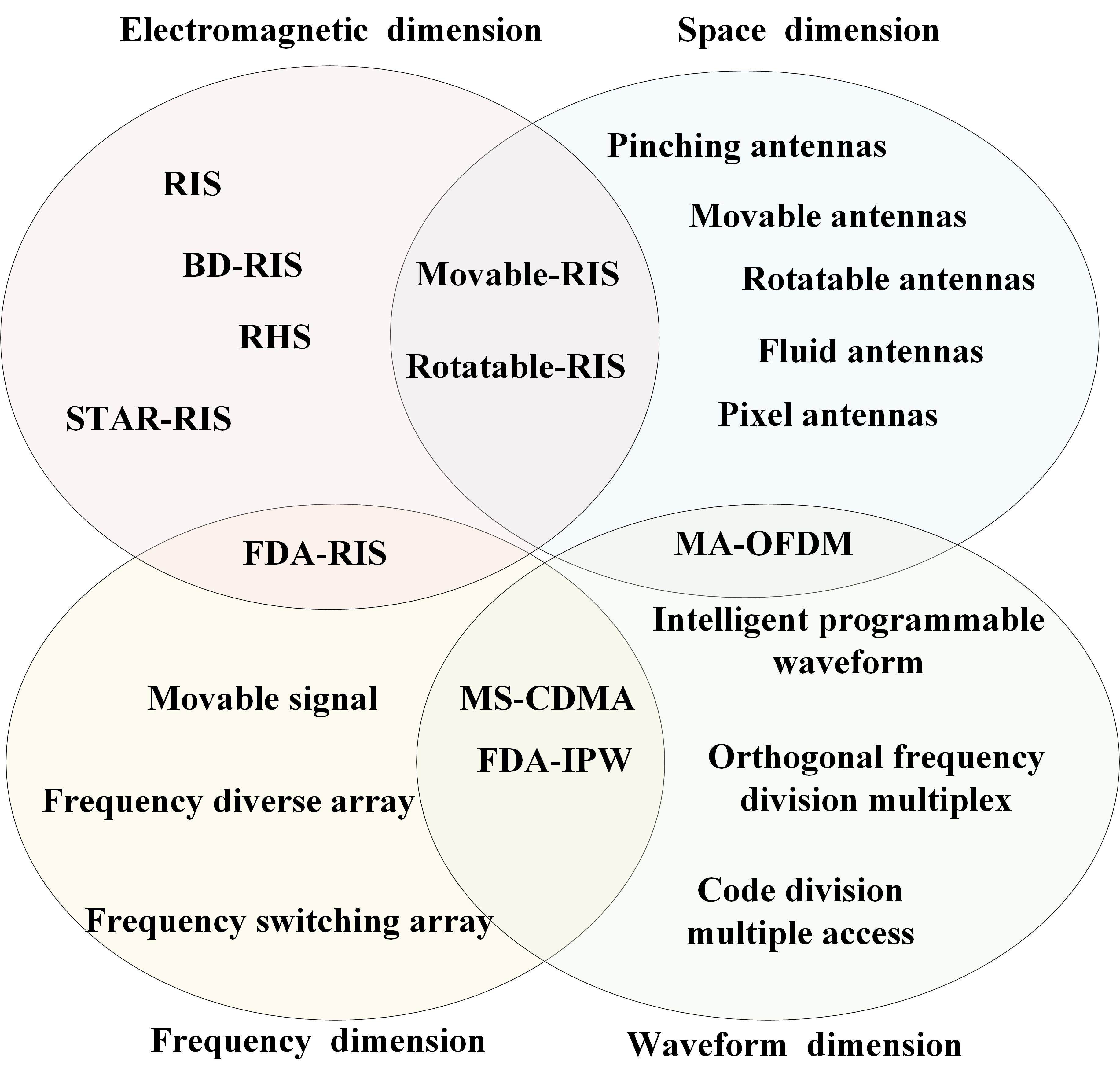}
\caption{Classification chart of SRE technologies in electromagnetic, space, frequency, and waveform dimensions.}
\label{Fig2}
\end{figure}

\subsection{SER Introduction}
In SRE, the wireless channel becomes an optimization variable. Active control over it can steer signal propagation and boost gain performance [6]. For dynamic, complex urban environments, this capability is a timely and powerful solution. Progress in electromagnetic information theory, advanced electromagnetics, and signal processing has spawned many promising solutions. These can be broadly divided into four categories, as illustrated in Fig.3.

\textbf{Electromagnetic dimension:} The electromagnetic dimension is a core fundamental element of wireless channels. Here, RIS stands out for its remarkable signal reconfiguration ability, which regulates the signal direction, phase, and amplitude [5]. As continuous advances, various derivative forms have emerged to overcome the shortcomings of conventional RIS, such as simultaneously transmitting and reflecting RIS (STAR-RIS) [11], reconfigurable holographic surface (RHS) [5], beyond-diagonal RIS (BD-RIS) [12], and so on.

\textbf{Space dimension:} The spatial dimension serves as the physical medium for wireless channel transmission. Movable antenna (MA) further exploits the degrees of freedom (DoFs) in space by flexibly adjusting the spacing between antenna elements, offering a new dimension for signal control. Common on antenna materials and actuation methods, typical MA types include six-dimension MA (6DMA), fluid antenna (FA), rotatable antenna (RA), pinching antenna (PCA), and pixel antenna, (PXA) [13]. Owing to their distinct traits, different MA types are suitable to different application scenarios, thereby providing various choices for LANs.

\textbf{Frequency dimension:} The frequency dimension is another essential face of wireless channels. Frequency diverse array (FDA) applies slight frequency offsets among antenna elements, allowing beam pattern manipulation and unlocking additional signal DoFs in the frequency dimension [4]. Moreover, movable signal (MS) and frequency switching arrays (FSAs) further exploit carrier frequency offsets to improve channel gain under limited spectrum resources [4]. These capabilities are critically significant for 6G systems, which face growing spectrum scarcity.

\textbf{Waveform dimension:} The waveform dimension is a key carrier for wireless channel transmission. Intelligent programmable waveform (IPW) actively adjusts the signal’s waveform structure to make its statistical properties resemble noise, thereby effectively hiding the signal and reducing the risk of detection. Furthermore, orthogonal frequency division multiplex (OFDM), spread spectrum, and code division multiple access (CDMA) all act within the waveform dimension. By optimizing the time-frequency structure and encoding schemes of signals, they significantly enhance the DoFs during signal propagation.

\textbf{Hybrid dimension:} With the rapid growth of low-altitude users, LANs serve diverse missions in complex electromagnetic environments. Relying on a single dimension often falls shorts, so multi-dimension cooperation and fusion are emerging as a critical trend for the evolution of LANs. Early research has already explored promising directions, such as movable RIS, FDA-RIS, and Rotatable RIS [14]. These technologies have shown promising improvements across LANs. As a result, multi-dimension cooperation and fusion are emerging as a prominent research focus going forward.

\subsection{Joint Design of SER and ISAC for LANs Covert Communication}

Leveraging wireless channel control capability of the SER can effectively enhance the covert performance and coverage of ISACC in complex low-altitude urban environments. Next, drawing on the distinct features of each SER dimension, we present four potential application scenarios tailored to four modes of ISACC in LANs, as illustrated in Fig. 3.

\begin{figure*}[!htb]   
\centering
\includegraphics[width=160mm,height=135mm]{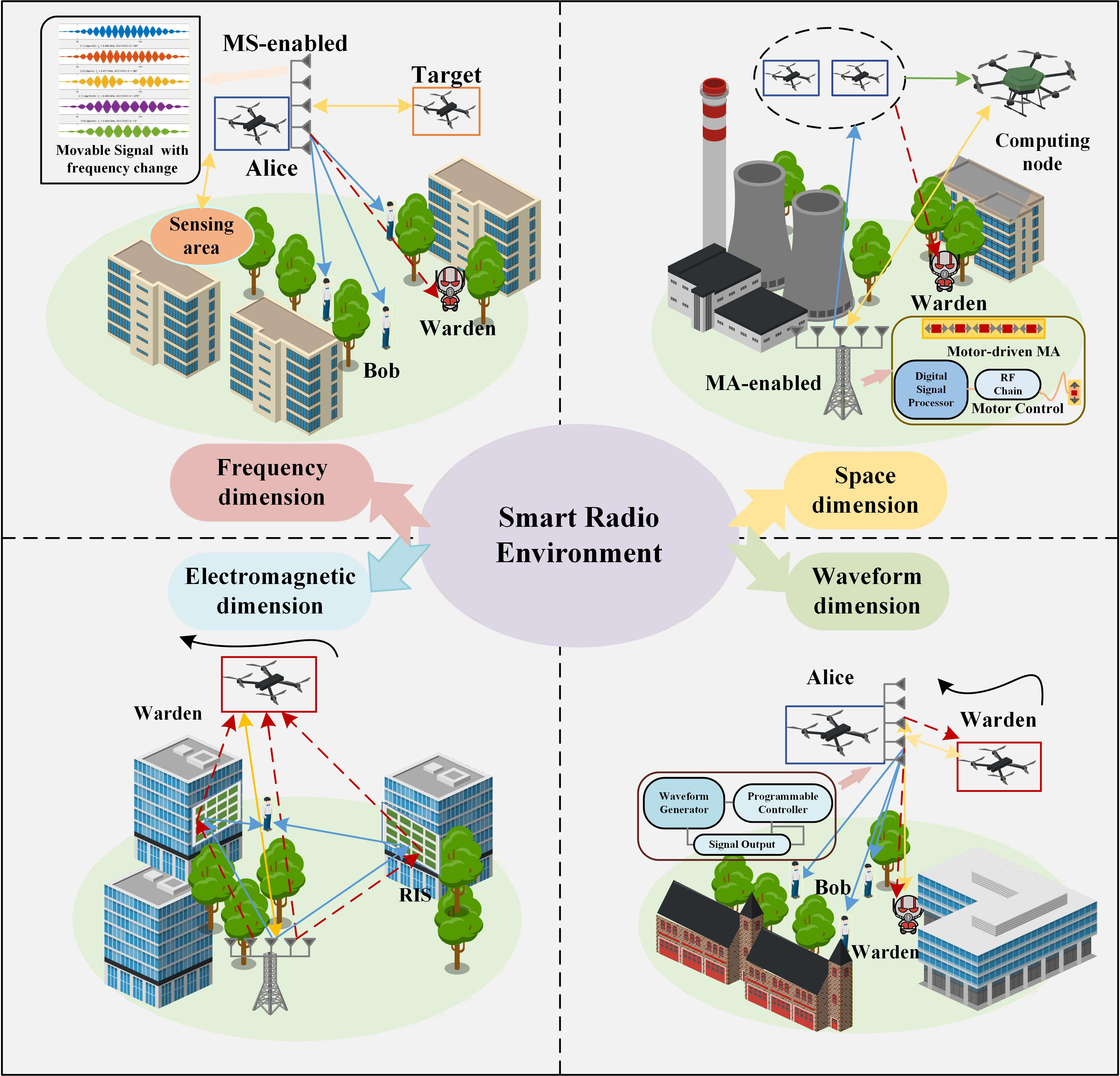}
\caption{The potential scenarios of SER-enabled ISACC in LANs: (a) MS-enabled ISACC in frequency dimension; (b) MA-enabled ISACC in space dimension; (c) RIS-enabled ISACC in electromagnetic dimension; (d) IPW-enabled ISACC in waveform dimension.}
\label{Fig1}
\end{figure*} 

\textbf{MS-CISAC-ABS:} In the frequency dimension, MS shifts the carrier frequency, which changes the signal's propagation angle and coverage. As shown in Fig. 3 (a), this not only reduces signal energy leakage toward the warden but also increases channel gain toward the user, thereby improving covert performance. Moreover, MS does not require phase shifters or tunable components, which reduces hardware complexity and energy consumption, making it more suitable for deployment on UAVs [4].

\textbf{MA-CISAC-BBS:} In the space dimension, MA moves individual antenna elements to shape both communication and sensing beams. As shown in Fig. 3 (b), MA-enabled GBS achieves precise target localization while simultaneously reducing signal energy leakage, thus enhancing covert transmission performance. Plus, incorporating MA modules into large-scale antenna arrays yields substantial performance improvements, making it advantageous for deploying large-scale array GBSs.

\textbf{RIS-NCISAC-BBS:} In the electromagnetic dimension, RIS can adjust the phase and amplitude of reflected signals by optimizing the individual elements. As shown in Fig. 3 (c),  RIS deployed on building walls not only redirects signals to alleviate the adverse effects of NLoS links on covert transmission, but also leverages sensing beams to monitor the warden's real-time position. This enables easier control of signal energy leakage toward the warden, thereby boosting covertness. Plus, due to the passive nature, RIS can be flexibly deployed in response to the unique characteristics of different urban environments.

\textbf{IPW-NCISAC-ABS:} In the waveform dimension, IPW can actively adjust the signal’s waveform to embed it into background noise. As shown in Fig. 3 (d), empowering IPW in ISAC could refine the integrated waveform of communication and sensing signals. This not only helps conceal signals into the noise but also enhances the energy efficiency of the system, making it attractive for energy constrained UAV platforms. In addition, refining the waveform structure and coding approach can boost channel gain, thereby delivering further covertness of ISACC in LANs.

\begin{figure*}[!htb]   
\centering
\includegraphics[height=16cm, width=19cm]{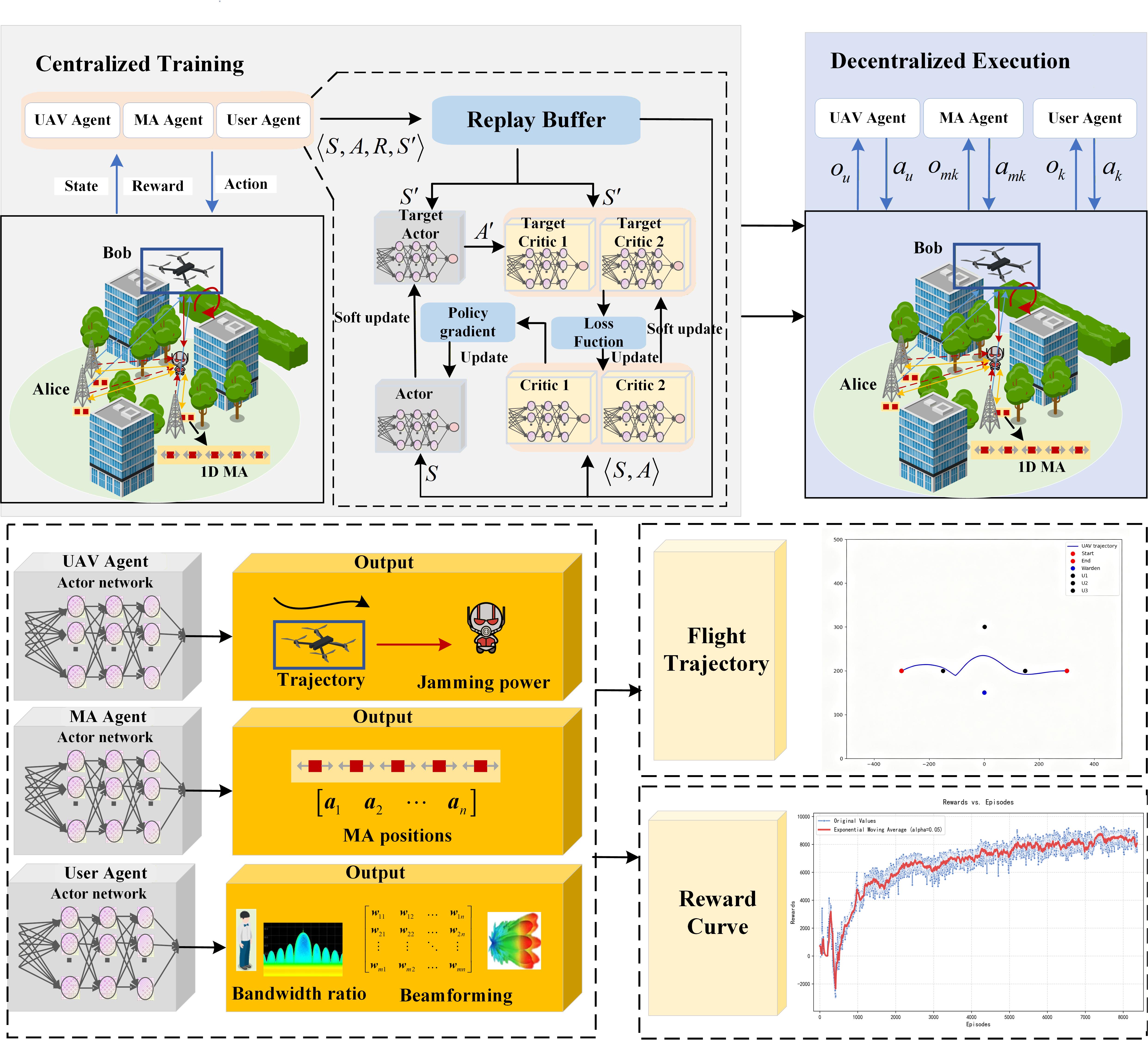}
\caption{The centralized training and decentralized execution framework of HMATD3 algorithm.}
\label{Fig1}
\end{figure*}         

\section{MA enabled UAV covert data collection in LANs }

When UAV performs the data collection task for GBSs, a terrestrial malicious node, acting as a warden, is capable of surveilling and monitoring the data transmission process. This framework locks down communication security from two sides. First, MA is deployed at the GBS-side to improve the SNR of the received signal. Second, the UAV transmits jamming signals to warden aimed at degrading its surveillance capabilities. Moreover, the GBSs emit sensing signals to achieve a specified gain at the warden location. To achieve reliable covert data collection strategy, a joint optimization of the UAV trajectory, jamming power, MA positions, bandwidth allocation, and the communication and sensing beamforming vectors is a critical procedure.

\subsection{System Model and Problem Formulation}
This section considers a UAV covert data collection system, where the UAV acted as an aerial receiver collecting data from K GBSs, and warden acted as an eavesdropper monitoring the data transmission. Moreover, GBSs also emit sensing signals to locate warden's position during covert transmission. Notably, due to the complexity of low altitude urban environment, the A2G links are modeled as channels incorporating both large-scale path loss and small-scale Rician fading. All GBSs employ the frequency division multiple access (FDMA) and are equipped with $N$ MAs. Furthermore, the UAV operates in full-duplex mode, simultaneously receiving the covert data and transmitting jamming signals towards warden to further disrupt his surveillance. Based on this, we design a multi-variable joint optimization framework that optimizes the UAV trajectory, jamming power, MA positions, bandwidth allocation, and the communication and sensing beamforming vectors to maximize the minimum covert rate across all GBSs.


First, according to the Rician fading model and the energy detection principle, we analyze warden's detection performance. Then, we derive the closed-form expression for the minimum detection error probability, transforming it into a tractable expression. Considering the highly coupled nature of variables and the complexity of the objective function in the original problem, we model it as a partially observable Markov decision process (PO-MDP). Building on the centralized training and decentralized execution (CTDE) framework, we proposed a heterogeneous multi-agent twin delayed deep deterministic policy gradient (HMATD3) algorithm for covert data collection. Specifically, as shown in Fig. 4, we model users, MA, and the UAV as three types of agents, and define their action spaces, state spaces, and reward functions according to the task requirements, respectively. During the training phase, the actor and critic networks are globally trained using data from the experience replay buffer. During the execution phase, each agent performs distributed policy execution leveraging these trained network parameters. Moreover, learning-based algorithm can complete pre-training via offline learning, which facilitates their practical deployment in time-sensitive scenarios. Therefore, compared to the conventional algorithms, HMATD3 not only achieves superior computational speed but also overcomes the impact of approximation errors inherent in conventional algorithms.

\subsection{Simulation Results}

As shown in Fig. 4, we consider a UAV covert data collection scenario with 3 GBSs and 1 warden, where the UAV's maximum velocity is $30 m/s$ and the covertness tolerance is set to $0.05$. In fight trajectory part of Fig. 4, the proposed algorithm’s reward curve reaches asymptotic convergence after 6000 iterations, verifying that a stable execution strategy is learned by agents to achieve the optimization objective of maximizing the minimum user covert rate. As shown in reward curve part of Fig. 4, the UAV first flies toward the centroid of the three GBSs for data collection, then approaches the warden to execute effective jamming. To balance the covert data collection rate and the effectiveness of jamming warden, the UAV maneuvers within the region between the warden and the GBSs. Finally, the UAV returns to its destination within mission duration.

In Fig. 5, the integration of MA provides a significant gain for UAV covert communication, particularly under relaxed covertness constraints. This is because, under such conditions, both UAV and the MAs have an expanded feasible region to select, thereby increasing the covertness gain. Notably, this advantage becomes more pronounced as the number of MAs increases, demonstrating that a larger MA array provides greater spatial DoFs for covert transmission, resulting in improving covert performance.

\begin{figure}[!htb]
\centering
\includegraphics[width=0.5\textwidth]{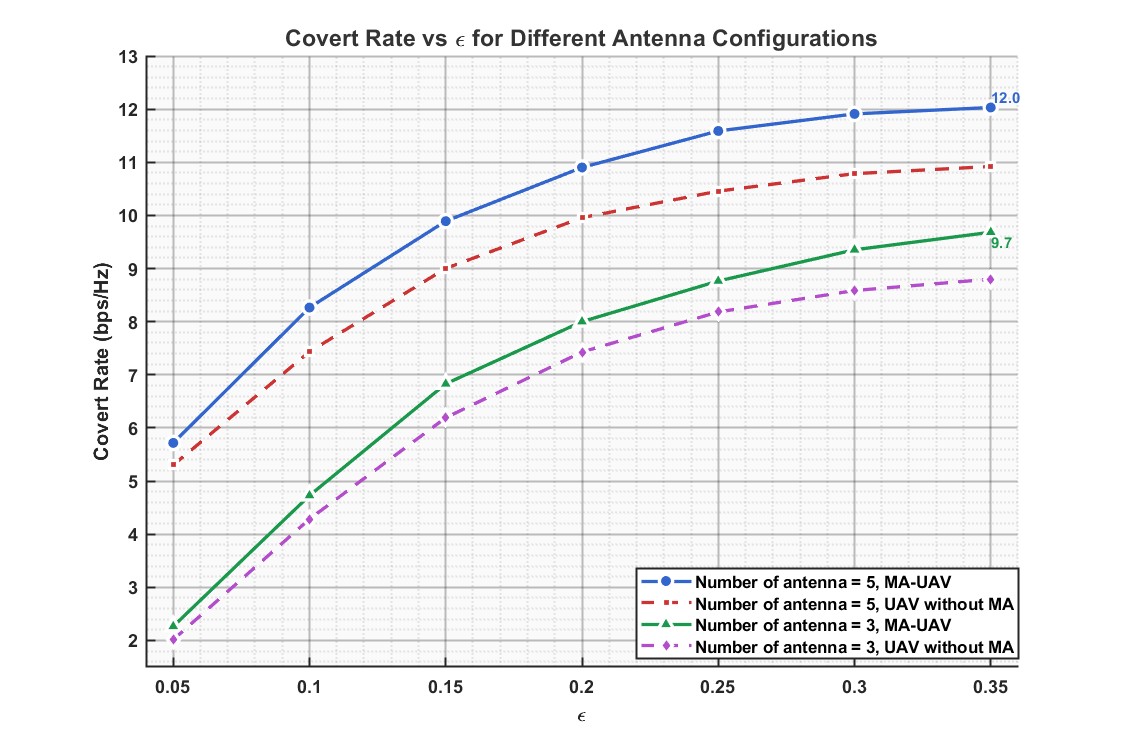}
\caption{The covert rate versus different covertness constraints $\varepsilon $.}
\label{Fig2}
\end{figure}

\section{Challenges and Future Directions}

Employing SRE technologies to enable ISACC is a novel and promising way to boost the covert communication in LANs. It must be emphasized that any technical means aimed at improving covert transmission in LANs must comply with applicable legal and regulatory frameworks, so as to deter malicious misuse by unauthorized or illegitimate entities. Furthermore, this paper identifies several interesting and challenging future research directions.

\textbf{Channel correlation for exploiting wavelength-scale mobility-induced uncertainty:} Due to the movement of MA elements in the space dimension, the wireless channels between each element exhibit strong spatial correlation. [15] employs a diagonal matrix approximation for spatial antenna correlation, developing a spatial block-correlation model. Nevertheless, no one has looked into how MA's channel correlation model affects the warden's surveillance. That makes channel correlation analysis a key stepping stone for unlocking MA's full covertness potential.

\textbf{Hybrid-dimension joint design for complex and dynamic LAN scenarios:} As the low-altitude economy advances, future LANs are growing more complex, exhibiting heterogeneity in nodes, dynamism in environments, and diversity in tasks. By employing SRE-enabled multi-dimension joint optimization, more DoFs can be unlocked to empower ISACC, thereby offering an effective pathway to address covert communication in LANs.

\textbf{Channel estimation and localization errors toward practical system design:} While SRE-enabled ISAC offers clear benefits for LACC, the integration of these inevitably raises signal processing complexity, which in turn makes channel estimation more challenging. Moreover, the precision of target sensing directly affects covertness. Therefore, quantifying the impact of channel estimation accuracy and localization errors on covertness is both necessary and urgent for practical systems.

\textbf{Empowered joint optimization for a dual-security approach for LANs:} While covert communication can address malicious signal detection in LANs, certain scenarios impose dual-security requirements, such as covertness and secrecy. That means fighting a warden detecting the communication activities, plus an eavesdropper attempting to decode the communication signals from physical layer security perspective. These scenarios present significant challenges, leveraging SRE to address them will become a highly intriguing and promising topic.

\section{Conclusion}

The integration of SRE with ISACC helps mitigate covert channel fading and improve the sensing precision in complex low-altitude urban environments. We began by reviewing the state of the art in ISACC and identifying key practical challenges. Then, we introduced the concept of SRE and detailed its enabling techniques across four dimensions: electromagnetic, frequency, spatial and waveform. Several promising integration scenarios were also discussed. Moreover, we proposed a novel scheme for LANs that introduces MA into ISACC could increase the space DoFs of channel and achieve superior covert throughput. Finally, some challenges and interesting direction for ISACC in LANs were pointed out.

\ifCLASSOPTIONcaptionsoff
  \newpage
\fi



%


\bibliographystyle{IEEEtran}

\section{Biographies}

{\textbf{{Jianyu Wei}}}
(wjianyu@aeu.edu.cn) received the B.S. degree in the Army Engineering University of PLA, Shijiangzhuang, China, in 2020, and the M.S. degree from the College of Communications Engineering, Army Engineering University of PLA, Nanjing, China, in 2022. He is currently pursuing the Ph.D. degree with the College of Communications Engineering, Army Engineering University of PLA, Nanjing, China. His research interests include covert communication, UAV communications, convex optimization techniques, and reinforcement learning.

{\textbf{Haichao Wang}}
(whcwl0919@sina.com) received the B.S. degree in electronic engineering and the Ph.D. degree in communications and information systems from the College of Communications Engineering, Army Engineering University of PLA, China, in 2014 and 2019, respectively. He is currently an associate professor with the College of Communications Engineering, Army Engineering University of PLA. His research interests include UAV communications, interference mitigation techniques, green communications, and convex optimization techniques.

{\textbf{Laixian Peng}}
(lxpeng@aeu.edu.cn) received the B.S. and Ph.D. degrees in telecom engineering from Nanjing Institute of Communications Engineering, Nanjing, China, in 1999 and 2004, respectively. Since 2008, he has been an Associate Professor with the PLA University of Science and Technology, Nanjing, where he was promoted as a Professor in 2016. His research interests include high-speed switching architectures and ad hoc networks and applications.

{\textbf{Jiangchun Gu}}
(gujiangchungjc@sina.com) received the B.S. degree in electronic and information engineering from Xidian University, Xi’an, China, in 2018, and the M.S. degree and Ph.D. degree in information and communication engineering from the College of Communications Engineering, Army Engineering University of PLA, Nanjing, China, in 2020 and 2023, respectively. He is currently a Lecturer with the College of Communications Engineering, Army Engineering University of PLA, China. His research interests include integrated communications and jamming, UAV communications, wireless communications, and convex optimization techniques.

{\textbf{Ziqi Liu}}
(dz1928005@smail.nju.edu.cn) received the Ph.D. degree in the School of Atmospheric Sciences, Nanjing University, Nanjing, China, in 2024, and currently works as an engineer at the Jiangsu Provincial Meteorological Observatory. Her research interests include numerical analysis, low-altitude meteorological communication network, and weather forecasting.

{\textbf{Lifeng Chen}}
(3011745933@qq.com) received the B.S. degree in 2022 from the College of Cryptography Engineering, People's Armed Police Force Engineering University, and the M.S. degree in 2024 from the same college. He is currently pursuing the Ph.D. degree with the College of Communications Engineering, Army Engineering University of PLA, Nanjing. His research interests include covert communication, UAV communications, edge computing, and reinforcement learning.

{\textbf{Guoru Ding}}
(guoru$\_$ding@yeah.net) is currently a Professor with the College of Communications Engineering, Army Engineering University of PLA, Nanjing, China. He is a recipient of a project supported by the National Science Fund for Distinguished Young Scholars of China. He has served as a Guest Editor for the IEEE JOURNAL ON SELECTED AREAS IN COMMUNICATIONS and an Associate Editor of the IEEE TRANSACTIONS ON COGNITIVE COMMUNICATIONS AND NETWORKING. His research interests include intelligent wireless communications and signal processing.

%








\end{document}